\documentclass[prb,reprint,superscriptaddress,a4paper,aps,showpacs,longbibliography]{revtex4-1}
\usepackage{graphicx}
\graphicspath{{./},{/home/user/fig/publi/muon/}}
\usepackage{amsfonts}
\usepackage{amsbsy}
\DeclareMathAlphabet\mathbfcal{OMS}{cmsy}{b}{n}

\begin{document}

\title{Determination of the zero-field magnetic structure of the helimagnet MnSi at low temperature}

\author{P. Dalmas de R\'eotier}
\affiliation{Universit\'e Grenoble Alpes, INAC-SPSMS, F-38000 Grenoble, France}
\affiliation{CEA, INAC-SPSMS, F-38000 Grenoble, France}
\author{A. Maisuradze}
\affiliation{Department of Physics, Tbilisi State University, Chavchavadze 3, 
GE-0128 Tbilisi, Georgia}
\author{A. Yaouanc}
\affiliation{Universit\'e Grenoble Alpes, INAC-SPSMS, F-38000 Grenoble, France}
\affiliation{CEA, INAC-SPSMS, F-38000 Grenoble, France}
\author{B. Roessli}
\affiliation{Laboratory for Neutron Scattering and Imaging, Paul Scherrer Institute, 5232 Villigen-PSI, Switzerland}
\author{A. Amato}
\affiliation{Laboratory for Muon-Spin Spectroscopy, Paul Scherrer Institute,
CH-5232 Villigen-PSI, Switzerland}
\author{D. Andreica}
\affiliation{Faculty of Physics, Babes-Bolyai University, 400084 Cluj-Napoca, Romania}
\author{G. Lapertot}
\affiliation{Universit\'e Grenoble Alpes, INAC-SPSMS, F-38000 Grenoble, France}
\affiliation{CEA, INAC-SPSMS, F-38000 Grenoble, France}

\date{\today}

\begin{abstract}

Below a temperature of approximately 29~K the manganese magnetic moments of the cubic binary compound MnSi order to a long-range incommensurate helical magnetic structure. Here, we quantitatively analyze a high-statistic zero-field muon spin rotation spectrum recorded in the magnetically ordered phase of MnSi by exploiting the result of representation theory as applied to the determination of magnetic structures. Instead of a gradual rotation of the magnetic moments when moving along a ${<}111{>}$ axis, we find that the angle of rotation between the moments of certain subsequent planes is essentially quenched. It is the magnetization of pairs of planes which rotates when moving along a ${<}111{>}$ axis, thus preserving the overall helical structure.

\end{abstract}


\maketitle

\section{Introduction}
\label{Intro} 

The determination of the magnetic structure of a magnet is usually performed using neutron diffraction through a two-step process.\cite{Rossat87} First the propagation wave vector of the structure is determined. Then a model for the Bragg reflection intensities leads to the magnetic structure. In some cases local probe techniques can help in the refinement of a magnetic structure previously inferred from neutron diffraction\cite{Pomjakushin00} or can even precede it.\cite{Vonlanthen99,Adroja15}

The MnSi compound has attracted attention since the first determination of its helical magnetic structure by neutron diffraction in the 1970s.\cite{Ishikawa76} Recently the detection of a skyrmion phase, first by neutron diffraction,\cite{Muhlbauer09a} has renewed the interest in this system.

In this paper we report a rather detailed analysis of a previously published field distribution measured by the muon spin rotation technique ($\mu$SR) at 5~K for a single crystal of MnSi.\cite{Amato14} This system is described as a helimagnet: while the Mn magnetic moments are ferromagnetically aligned in planes perpendicular to the magnetic propagation vector ${\bf k} \parallel {<}111{>}$, the direction of the moments in subsequent planes slightly rotates with a phase merely given by the scalar product ${\bf k}\cdot{\bf r}$, ${\bf r}$ denoting the position of a Mn atom in the crystal. Here we show that an additional phase is present that distinguishes the Mn atoms for which the local symmetry axis is or is not collinear to ${\bf k}$. As a consequence of this phase shift, the angle between moments in some subsequent planes nearly vanishes, i.e., they are in fact quasi-ferromagnetically aligned.

We shall first present some physical properties of MnSi in Sec.~\ref{MnSi_basic}. This will allow us to introduce some notations to be used later on. In the next section (Section~\ref{Muon_Amato}) we shall summarize the recently published results obtained from a $\mu$SR study of MnSi.\cite{Amato14} Section~\ref{Symmetry} examines the magnetic structures of MnSi compatible with its crystallographic symmetry. In Sec.~\ref{New_result}, using a method powerful enough to account for a local deviation of the magnetic structure from the well-known slowly varying magnetic density, we describe the zero-field (ZF) $\mu$SR spectrum recorded at 5~K for MnSi. A discussion of our result which is compatible with published neutron diffraction results is given in Sec.~\ref{Discussion}. Some conclusions are gathered in Sec.~\ref{Conclusions}. In Appendix~\ref{Group_theory_application}, using representation theory as applied to the selection of magnetic structures consistent with crystalline symmetries, we determine the possible solutions for MnSi.  Appendix~\ref{Group_theory_basic} lists some basic results from representation theory.

\section{Some physical properties of M\lowercase{n}S\lowercase{i}} 
\label{MnSi_basic}

The room temperature crystal structure of the compound MnSi was solved by Bor\'en already in 1933.\cite{Boren33} It crystallizes in the cubic P2$_1$3 space group with the so-called B20 structure, a structure with no center of symmetry. The lattice parameter is $a_{\rm lat}$ = 4.558~\AA. The Mn and Si atoms occupy $4a$ Wyckoff positions. The coordinates of a $4a$ position depend on a single parameter $x$, i.e., $x_{\rm Mn}$ = 0.138 and $x_{\rm Si}$ = 0.845. Namely, the four equivalent positions are $(x,x,x)$, $(\bar{x}+\frac{1}{2}, \bar{x}, x+\frac{1}{2})$, $(\bar{x}, x+\frac{1}{2}, \bar{x}+\frac{1}{2})$, and $(x+\frac{1}{2},\bar{x}+\frac{1}{2},\bar{x})$. 

The MnSi structure can be viewed as a stacking of Mn and Si planes perpendicular to a three-fold axis, say [111]. Hereafter we will only consider the Mn sites. Two types of alternatively stacked Mn planes can be distinguished. One is constituted by atoms whose local trigonal axis is collinear to [111]. The other type of planes contains the other atoms with a local trigonal axis symmetry-equivalent to, but different from [111]. The atomic population of the two types of planes is in the ratio 1:3. Interestingly the planes are not equidistant: two interplanar distances are found, 1.179 and 1.453~\AA.

The compound exhibits a magnetic transition near 29~K with a magnetic moment of $m = 0.4 \, \mu_{\rm B}$ per Mn atom at low temperature.\cite{Williams66} The first order nature of the transition has slowly emerged, with some theoretical evidence already presented in 1980.\cite{Bak80} From neutron diffraction it has been established that the Mn magnetic moments form a left-handed helix with an incommensurate propagation vector ${\bf k}_\ell$ parallel to one of the four three-fold axis.\cite{Ishikawa76,Ishida85} Here the subscript $\ell$ labels one of the four K-domains as discussed below. Since $k_\ell \equiv k \simeq 0.35$~nm$^{-1}$ at low temperature, the helix period is $2\pi/k = \lambda_{\rm h} \simeq$~18~nm. In the assumed structure, the Mn magnetic moments are ferromagnetically aligned in planes perpendicular to ${\bf k}_\ell$ while, from one plane to the following, their orientation slightly rotates. The mathematical expression for the magnetic moment at position ${\bf r}$ is given by the formula
\begin{equation}
{\bf m}_\ell({\bf r}) = m [ \cos({\bf k}_\ell\cdot{\bf r} + \alpha_0) {\bf a}_\ell - 
\sin({\bf k}_\ell\cdot{\bf r} + \alpha_0) {\bf b}_\ell].
\label{moment_neutron}
\end{equation}
Here $\alpha_0$ is a phase and ${\bf a}_\ell$ and ${\bf b}_\ell$ are two orthonormal vectors orthogonal to ${\bf k}_\ell$, with ${\bf k}_\ell/k$ = ${\bf a}_\ell \times {\bf b}_\ell$.

The system is characterized by a hierarchy of three energy scales comprising a dominant ferromagnetic symmetric exchange interaction, a relatively weak antisymmetric Dzyaloshinski-Moriya (DM) exchange interaction caused by the lack of a center of symmetry in the crystal structure and a much weaker anisotropic exchange interaction fixing the direction of propagation of the magnetic spiral along one of the cube diagonals. Taking into account the two exchange terms and following Ginzburg and Landau, the free energy functional can be expanded in terms of a slowly varying magnetic density consistent with Eq.~\ref{moment_neutron}.\cite{Bak80,Nakanishi80}

Two types of magnetic domains exist. To the four branches of the star formed by the magnetic structure propagation vectors correspond four K-domains, i.e., ${\bf k}_\ell$ can be collinear to $[111]$, $[\bar{1}\bar{1}1]$, $[\bar{1}1\bar{1}]$, or $[1\bar{1}\bar{1}]$. We will denote these domains by the letters A, B, C, and D, respectively, i.e., $\ell \in \{{\rm A},{\rm B}, {\rm C}, {\rm D} \}$. 
We also need to pay attention to the three spin-domains: i.e., for a given K-domain there are three S-domains corresponding to a phase equal to $\alpha_0$, $\alpha_0 +2 \pi/3$ and $\alpha_0 +4 \pi/3$ in Eq.~\ref{moment_neutron}, respectively. However, since we are interested in the field distribution measured by $\mu$SR we shall integrate over a phase $\zeta$ varying between $0$ and $2 \pi$ to account for the incommensurate structure; see Sect.~\ref{New_result} for full details. This implies that the existence of the S-domains has no influence on the computed $\mu$SR field distribution.

\section{Summary of recent $\mu$SR results}
\label{Muon_Amato}

Two main experimental results were obtained from a recent $\mu$SR study of single crystalline MnSi.\cite{Amato14}

The angular dependence of the muon precession frequencies measured with a transverse field in the paramagnetic phase matches expectation for the muon in a $4a$ position with $x_{\mu^+}$ = 0.532. Hence, to a crystallographic $4a$ position correspond four different muon magnetic sites, that we will identify with the index $\eta$, namely $\eta\in\{1,2,3,4\}$. The aforementioned angular dependence reflects the symmetry of the dipole field acting on the muon spin and arising from the lattice of Mn moments. In addition to this field, an isotropic hyperfine contact field, described by a constant to be defined in Sec.~\ref{New_result}, was measured. In the following we will denote ${\bf r}_{0,s_\eta}$ the vector distance between the muon site $s_\eta$ and the origin of the cubic lattice. Remarkably, there is no need to invoke a muon-induced effect to understand the paramagnetic data. 

Building on the determination of the crystallographic muon position, the field distribution derived from a Fourier transform of a high statistic ZF spectrum at 5~K was discussed in terms of the distribution of fields expected for the magnetic density of Eq.~\ref{moment_neutron}.  It was argued that muons in three out of the four muon magnetic sites sense identical field distributions typical of an incommensurate helical magnetic structure. These are continuous field distributions characterized by a lower and a upper cut-off fields.\cite{Schenck01} On the other hand, muons at the remaining site probe a unique third field irrespective of the phase $\alpha_0$ of the helix. The shape of the distribution deduced from the Fourier transform of the asymmetry spectrum supports this analysis. However, the difference between the lower cutoff and third fields was predicted to be 2~mT whereas a splitting of 4.8\,(4)~mT was measured.\cite{Amato14} This discrepancy cannot be resolved by tuning the values of $x_{\mu^+}$ or of the hyperfine contact field. Surprisingly, we found that it can be resolved assuming $k$ to be approximately twice as large as expected from neutron diffraction.\cite{Ishikawa76} This cannot be correct and therefore suggests the description of the magnetic density as expressed by Eq.~\ref{moment_neutron} to be a too rough approximation for the description of the ZF $\mu$SR spectrum.

\section{Symmetry analysis of the magnetic structure of M\lowercase{n}S\lowercase{i}}
\label{Symmetry}

\begin{figure}
\centering
\includegraphics[width=0.25\textwidth]{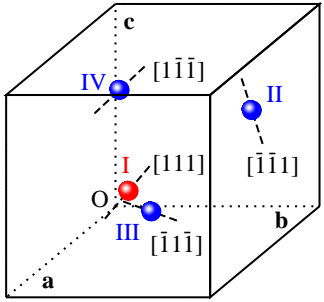}
\caption{(Color online) Position of the Mn atoms in the crystal structure of MnSi. The three edges ${\bf a}$, ${\bf b}$, and ${\bf c}$ of the unit cubic cell of origin O are indicated. For each position, the dashed line represents the direction of the local 3-fold symmetry axis indicated nearby. Considering domain A with propagation wavevector ${\bf k}_{\rm A}$ parallel to [111], the local 3-fold symmetry axis is parallel to ${\bf k}_{\rm A}$ only for Mn site I. This explains why site I on the one hand and sites II, III, and IV on the other hand belong to two different orbits.}
\label{crystal_structure}
\end{figure}

The magnetic density in Eq.~\ref{moment_neutron} was obtained from a Ginzburg-Landau expansion valid at a mesoscopic scale. At a microscopic level it may be worth paying attention to each of the four Mn atoms in a unit cell. We specify the position of a unit cell by the cubic lattice vector $ {\bf i }$ and that of a Mn atom within a cell by $ {\bf d}_\gamma$ with $\gamma \in \{{\rm I},{\rm II},{\rm III},{\rm IV} \}$. For a magnetic moment at position $ {\bf i + d_\gamma}$ we write 
\begin{eqnarray}
{\bf m}_{\ell,i + d_\gamma} & = & m \left \{  \cos \left[{\bf k}_\ell\cdot ({\bf i + d_\gamma}) + \alpha_{\ell,d_\gamma} \right] {\bf a}_{\ell,d_\gamma}  \right . \cr
& &  - \left . \sin \left [{\bf k}_\ell\cdot ({\bf i + d_\gamma}) + \alpha_{\ell,d_\gamma} \right ] {\bf b}_{\ell,d_\gamma} \right \}.
\label{moment_muon_general}
\end{eqnarray}
Here we still assume helices, i.e., ${\bf a}_{\ell,d_\gamma}$ and ${\bf b}_{\ell,d_\gamma}$ are orthogonal unit vectors, but we recognize that they may be different for the four Mn atoms in a cell. In fact, we are only looking for a small deviation from the magnetic structure predicted by  Eq.~\ref{moment_neutron}. To proceed further we need to determine the magnetic structures compatible with the crystallographic symmetry. This is done in Appendix~\ref{Group_theory_application} where it is shown that two kinds --- the so-called orbits --- of Mn atoms must be considered; see also Fig.~\ref{crystal_structure}. From this symmetry analysis we basically get two results. (i) A phase shift $\phi$ between the moments in the two orbits may exist, the value of which is not provided by the analysis. (ii) For a given k-domain the plane defined by ${\bf a}_{\ell,d_\gamma}$ and ${\bf b}_{\ell,d_\gamma}$ may not be orthogonal to ${\bf k}_\ell$. The rotation axis ${\bf n}_{\ell,d_\gamma} = {\bf a}_{\ell,d_\gamma} \times {\bf b}_{\ell,d_\gamma}$ is defined by its polar $\theta$ and azimuthal $\varphi$ angles in the MnSi cubic frame. Hence, compared to the magnetic structure given by Eq.~\ref{moment_neutron} we have three additional free parameters. In the fitting procedure we will allow them to deviate from the original values $\phi_0$ = 0, $\theta_0$ = 54.7$^\circ$ and $\varphi_0$ = 45$^\circ$ for which the structure corresponds to Eq.~\ref{moment_neutron}. In fact we will find that the experimental data are well accounted for with $\theta = \theta_0$ and $\varphi = \varphi_0$, i.e., we can choose ${\bf a}_{\ell,d_\gamma}$ and ${\bf b}_{\ell,d_\gamma}$ independent of $\gamma$, and three of the four $\alpha_{\ell,d_\gamma}$ phases to be equal. We shall set them to 0 without loss of generality. Only a relative change $\phi$ of the fourth phase is necessary to obtain a proper fit.

\section{Analysis of the ZF $\mu$SR spectrum of M\lowercase{n}S\lowercase{i} at 5~K based on symmetry}
\label{New_result}

Before proceeding to the analysis of the ZF $\mu$SR spectrum, we expose a few experimental details. The crystal used for this experiment was grown by the Czochralski method and already served for $\mu$SR experiments reported in Ref.~\onlinecite{Yaouanc05}. The ZF spectrum of interest in the present study was recorded with the GPS spectrometer of the Swiss Muon Source (Paul Scherrer Institute, Villigen, Switzerland). The [111] crystal axis was set collinear to the muon initial polarization.

The standard ZF setup was used for this experiment.\cite{Yaouanc11} Polarized muons are implanted in the sample under study and the evolution of the muon spins is monitored through the anistropic decay of the muon. The decay positrons are detected in two counters set parallel and antiparallel to the muon initial polarization. Conventionally, the Cartesian axis defining this initial polarization is labeled $Z$. 

\subsection{The polarization function}\label{Analysis_pol_func}

Solving the Larmor equation for a muon spin ${\bf S}$ submitted to a magnetic field ${\bf B}$, the evolution of the spin $Z$-component is readily obtained:
\begin{eqnarray}
\frac{S_Z(t)}{S} & = & \left(\frac{B^Z}{B}\right)^2 + \left[1-\left(\frac{B^Z}{B}\right)^2\right] \cos(\gamma_\mu B t).
\label{P_Z-1}
\end{eqnarray}
Here $\gamma_\mu$ = 851.6~Mrad\,s\,T$^{-1}$ is the muon gyromagnetic ratio.

Considering an experiment in the ordered phase of MnSi, the spontaneous field ${\bf B}_{0,\ell,s_\eta}$ to which the muon at site $s_\eta$ in magnetic domain $\ell$ is submitted is the vectorial sum of the dipole field arising from the Mn magnetic moments and a contact field that we will assume isotropic. The quantity $S_Z(t)$ is obviously given by Eq.~\ref{P_Z-1} after setting ${\bf B}$ = ${\bf B}_{0,\ell,s_\eta}$.

Since ${\bf B}_{0,\ell,s_\eta}$ linearly depends on the Mn magnetic moments it can be expressed in terms of a product of a coupling tensor $\boldsymbol{G}$ with a magnetic moment vector. For reasons that will become clear later, it is of interest to recast to the Fourier components of the magnetic moments. Following Ref.~\onlinecite{Yaouanc11} we write \footnote{Equations~\ref{spontaneous_field} and \ref{Moment_Fourier} derive from Eqs.~5.60 and 5.61 of Ref.~\onlinecite{Yaouanc11}}
\begin{eqnarray}
{\bf B}_{0,\ell,s_\eta} & = & {\mu_0 \over 4 \pi} {1 \over \sqrt{n_{\rm c}} v_{\rm c}}
\label{spontaneous_field}\\
& & \times 
\sum_{\gamma} \sum_{{\bf q}} {\boldsymbol G}_{d_\gamma, {\bf q},s_\eta} {\bf m}_{\ell,d_\gamma, {\bf q}} 
\exp (-i {\bf q} \cdot {\bf r}_{0,s_\eta} ),\nonumber
\end{eqnarray}
where 
%
\begin{eqnarray}
{\bf m}_{\ell,d_\gamma, {\bf q}}  = {1 \over \sqrt n_{\rm c}} \sum_{\bf i} {\bf m}_{\ell, i + d_\gamma} 
\exp \left [-i {\bf q} \cdot \left({\bf i + d_\gamma} \right) \right]
\label{Moment_Fourier}
\end{eqnarray}
is the Fourier component of the sublattice magnetic moment. Here $\mu_0$ is the permeability of free space, $n_{\rm c}$ is the number of unit cells in the crystal under study and $v_{\rm c}$ = $a_{\rm lat}^3$ is their volume. The sum in Eq.~\ref{spontaneous_field} is performed over the ${\bf q}$ vectors of the first Brillouin zone (BZ).

Reflecting the two contributions to ${\bf B}_{0,\ell,s_\eta}$ we write ${\boldsymbol G}_{d_\gamma, {\bf q},s_\eta}$ as the sum of the dipolar and contact field terms:
\begin{eqnarray}
{\boldsymbol G}_{d_\gamma, {\bf q},s_\eta} & = & {\boldsymbol D}_{d_\gamma, {\bf q},s_\eta} + {\boldsymbol H}_{d_\gamma, {\bf q},s_\eta}.
\label{G=D+H}
\end{eqnarray}
The dipolar term ${\boldsymbol D}_{d_\gamma, {\bf q},s_\eta}$ is computed following the Ewald summation technique.\cite{Yaouanc93,Yaouanc93a,Born54} This method ensures a fast and exact evaluation of the lattice sum which is slowly converging. 
The element of Cartesian components $\beta_1\beta_2$ of the contact field tensor is written
\begin{eqnarray}
H_{d_\gamma, {\bf q},s_\eta}^{\beta_1\beta_2} & = & \sum_{\bf i}{}^\prime n_d H \delta_{\beta_1\beta_2} \exp\left[i{\bf q}\cdot({\bf i} + {\bf d}_\gamma+{\bf r}_{0,s_\eta})\right],
\label{Contact_field}
\end{eqnarray}
where $n_d = 4$ is the number of Mn sublattices in the cubic crystal structure. The assumed isotropy of the contact field interaction is enforced by the Kronecker symbol $\delta_{\beta_1\beta_2}$. Recognizing the short range character of this interaction the sum is limited to the first nearest neighbors to the muon. In fact, in the MnSi crystal structure there are $r_\mu = 3$ nearest neighbors to the muon belonging to different sublattices. Therefore, for a given muon site and a $\gamma$ sublattice the sum in Eq.~\ref{Contact_field} is limited to 1 or even 0 term. In the numerical investigation to be presented below we have performed tests including the second nearest neighbors. No significant change to our results has been found.
From the study performed in the paramagnetic phase,\cite{Amato14} and the value 
$A_{\rm cont,TF} = -0.9276\,(20)$~mol/emu derived with different notations for the contact field coupling, we obtain $r_\mu H/4\pi$ = $-1.052\,(2)$ for the relevant parameter describing the contact field.\cite{Yaouanc11} This value will be our starting value for the fit of the spectrum in the ordered phase.

For future reference we provide an expression for ${\bf m}_{\ell,d_\gamma, {\bf q}}$. Introducing ${\bf m}_{\ell, i + d_\gamma}$ given by Eq.~\ref{moment_muon_general} into Eq.~\ref{Moment_Fourier} and using the equality $\sum_{\bf i} \exp[i({\bf q}-{\bf q}')\cdot {\bf i}] = n_{\rm c}\delta_{{\bf q},{\bf q}'}$ where $\delta_{{\bf q},{\bf q}^\prime}$ is a Kronecker symbol, we derive 
\begin{eqnarray}
{\bf m}_{\ell,d_\gamma, {\bf q}} & = & \delta_{{\bf q},\pm{\bf k}_\ell} {\bf m}_{\ell,d_\gamma, \pm} 
\label{Moment_Fourier_component}
\end{eqnarray}
with
\begin{eqnarray}
{\bf m}_{\ell,d_\gamma, \pm}  & = & \frac{\sqrt{n_{\rm c}} m}{2}
\exp(\pm i \alpha_{\ell,d_\gamma} ) \left( {\bf a}_{\ell,d_\gamma}
\pm i {\bf b}_{\ell,d_\gamma}\right).
\label{Moment_Fourier_component_expression}
\end{eqnarray}

For a typical $\mu$SR experiment millions of muons are implanted in the compound under study. Considering the muons localized at site $s_\eta$, the set of vector positions is $\{({\bf r}_{0,s_\eta}$ + ${\bf i}), {\bf i}\in {\rm DL}\}$ where ${\bf i}$ is a vector of the cubic direct lattice (DL). Owing to the incommensurate nature of the magnetic order the set of values spanned by the factor $\exp[\mp i{\bf k}_\ell\cdot({\bf r}_{0,s_\eta}+{\bf i})]$ entering in Eq.~\ref{spontaneous_field} is $\{\exp[\mp i({\bf k}_\ell\cdot {\bf r}_{0,s_\eta}+\zeta)],\ \zeta \in [0, 2\pi[\}$.\cite{Schenck01,Yaouanc11} Here we have recognized that ${\bf m}_{\ell,d_\gamma,{\bf q}}$ vanishes unless ${\bf q} = \pm {\bf k}_\ell$ (Eq.~\ref{Moment_Fourier_component}). The distribution of fields at the muon sites is therefore obtained after an integration over $\zeta$:
\begin{eqnarray}
D_{\rm v}({\bf B}) = \frac{1}{2\pi}\int_{0}^{2\pi}  \delta[{\bf B} 
& - &\boldsymbol{\mathfrak{B}}_{0,\ell,{\bf k}_\ell,{\rm s}_\eta}(-{\bf k}_\ell\cdot {\bf r}_{0,s_\eta}-\zeta) \label{distribution_alpha}\\
& - & 
\boldsymbol{\mathfrak{B}}_{0,\ell,-{\bf k}_\ell,{\rm s}_\eta}({\bf k}_\ell\cdot {\bf r}_{0,s_\eta}+\zeta)
] \,{\rm d}\zeta,\nonumber
\end{eqnarray}
where the expression for 
\begin{eqnarray}
\boldsymbol{\mathfrak{B}}_{0,\ell,{\bf q},s_\eta}(\psi) & = & {\mu_0 \over 4 \pi} \frac{1}{\sqrt{n_{\rm c}} v_{\rm c}} 
\sum_{\gamma} 
{\boldsymbol G}_{d_\gamma, {\bf q},s_\eta} {\bf m}_{\ell,d_\gamma, {\bf q}}
\exp (i \psi ),\cr & &
\label{spontaneous_field_int}
\end{eqnarray}
is derived from Eq.~\ref{spontaneous_field}. The polarization function associated with muons stopped at site $s_\eta$ in domain $\ell$ reads
\begin{eqnarray}
P_{Z,\ell,s_\eta}(t) = \int \frac{S_Z(t)}{S} D_{\rm v}({\bf B})\, {\rm d}^3{\bf B}.
\label{Polarization_l_eta}
\end{eqnarray}
For an expression of the muon polarization function $P_Z(t)$ we must average over the four muon sites and over the four k-domains, i.e.\ 
\begin{eqnarray}
P_Z(t) = \langle P_{Z,\ell,s_\eta}(t) \rangle_{\ell,\eta}.
\label{Fonction_Polarization}
\end{eqnarray}

Before considering the experimental data we need to include the effect of three physical phenomena which have not been addressed in the model described so far. The muon spin-lattice relaxation evidenced in previous experiments (see e.g. Refs.~\onlinecite{Takigawa80,Yaouanc05}) is not accounted for in our model. For this purpose we will phenomenologically include an $\exp(-\lambda_Zt)$ factor to the first term in the right-hand side of Eq.~\ref{P_Z-1}, where $\lambda_Z$ is the spin-lattice relaxation rate. The other phenomena to be addressed concern sources of damping of the oscillations associated with the second term in Eq.~\ref{P_Z-1}. Besides the magnetic field of electronic origin described by Eq.~\ref{spontaneous_field}, the muons are sensitive to the field produced by the nuclei of the compound. Assuming the components of this field to be Gaussian distributed with a root-mean-square $\Delta_{\rm N}$, we add an $\exp(-\gamma_\mu^2\Delta_{\rm N}^2 t^2/2)$ factor to the second term in Eq.~\ref{P_Z-1}. Finally the coherence length of the magnetic structure is not infinite. In diffraction experiments this leads to Bragg spots with an extension which can overpass the width given by the diffractometer resolution. In $\mu$SR this contributes to a further damping of the oscillations. To account for this effect we replace the discrete sum over BZ in Eq.~\ref{spontaneous_field} by an integral and the Kronecker symbol in Eq.~\ref{Moment_Fourier_component} by a continuous function. Altogether we perform the substitution
\begin{eqnarray}
\frac{S_Z(t)}{S} & \longrightarrow & \left(\frac{B^Z}{B}\right)^2 \exp(-\lambda_Z t)
\label{P_Z-2}\\
& + & \left[1-\left(\frac{B^Z}{B}\right)^2\right] \exp\left(-\frac{\gamma_\mu^2\Delta_{\rm N}^2 t^2}{2}\right) \cos(\gamma_\mu B t).\cr
\nonumber
\end{eqnarray}
in Eq.~\ref{Polarization_l_eta} and replace Eq.~\ref{distribution_alpha} with
\begin{eqnarray}
D_{\rm v}({\bf B}) & = & \frac{v_{\rm c}}{(2\pi)^4}
\label{distribution_alpha_eta}\\
& \times & 
\int_{\rm BZ}\int_0^{2\pi}  \delta[{\bf B} - \boldsymbol{\mathfrak{B}}_{0,\ell,{\bf q},{\rm s}_\eta}(-{\bf q}\cdot {\bf r}_{0,s_\eta}-\zeta)] {\rm d}\zeta\,{\rm d}^3 {\bf q}.
\nonumber
\end{eqnarray}
In addition we make the following change in Eq.~\ref{spontaneous_field_int}:
\begin{eqnarray}
{\bf m}_{\ell,d_\gamma, {\bf q}} & \longrightarrow & f_{\ell,d_\gamma}({\bf q} -{\bf k}_\ell) {\bf m}_{\ell,d_\gamma,+} + f_{\ell,d_\gamma}({\bf q} +{\bf k}_\ell) {\bf m}_{\ell,d_\gamma,-}.\cr & &
\label{M_q}
\end{eqnarray}
Here we have defined
\begin{eqnarray}
f_{\ell,d_\gamma}^2({\bf q}) & \propto & \frac{\xi^3}{1+\xi^2q^2}
\label{f_q}
\end{eqnarray}
where $\xi$ is the coherence length of the magnetic structure. This form for $f_{\ell,d_\gamma}({\bf q})$ corresponds to a Lorentzian lineshape for the intensity recorded around a Bragg position in a scattering measurement.\cite{deGennes60} To ensure convergence of the integral in Eq~\ref{distribution_alpha_eta}, we limit the integration to $-5  \le \xi q_\beta \le 5$ for each Cartesian component $\beta$ of ${\bf q}$ and compute the proportionality constant in Eq.~\ref{f_q} accordingly. 
As before $P_Z(t)$ is computed from $P_{Z,\ell,s_\eta}(t)$ using Eq.~\ref{Fonction_Polarization}.

\subsection{Results}\label{results}

\begin{figure}
\centering
\includegraphics[width=0.45\textwidth]{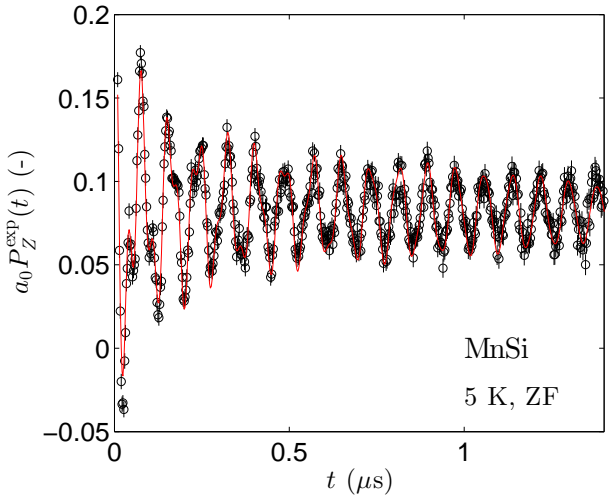}
\caption{(Color online) 
A ZF $\mu$SR asymmetry spectrum recorded at 5~K with the $Z$ axis parallel to a three-fold axis of a crystal of MnSi. The solid line is the result of a fit discussed in the main text. }
\label{Spectrum}
\end{figure}

\begin{figure}
\centering
\includegraphics[width=0.45\textwidth]{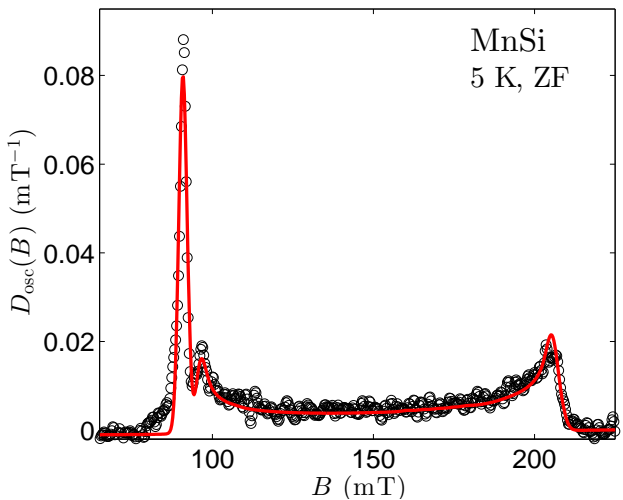}\\
\caption{(Color online) 
Field distribution associated with the oscillating part of the asymmetry of Fig.~\ref{Spectrum}. The solid line results from a fit explained in the main text. 
}
\label{Fourier}
\end{figure}

A $\mu$SR asymmetry spectrum $a_0 P_Z^{\rm exp}(t)$ recorded on a crystal of MnSi and originally presented in Ref.~\onlinecite{Amato14} is displayed in Fig.~\ref{Spectrum}. The so-called initial asymmetry $a_0$ is an instrumental parameter.
While only $a_0 P_Z^{\rm exp}(t)$ is fitted, it is useful to consider the field distribution associated with its oscillating part, namely $D_{\rm osc}(B)$. It is pictured in Fig.~\ref{Fourier}.
Three characteristic fields are found: 91~mT for the isolated peak and 96 and 207~mT respectively for the lower and upper cutoffs of the continuous part of the distribution.\footnote{Only two spontaneous fields have been observed for asymmetry spectra recorded at lower statistics,\cite{Kadono90,Uemura07,Andreica10} i.e. the field at 96~mT is absent} The solid line in Fig.~\ref{Spectrum} results from a fit of the model described in Sect.~\ref{Analysis_pol_func}, with $P_Z^{\rm exp}(t) = P_Z(t)$ and the magnetic structure given in Appendix~\ref{Group_theory_application}, to the data. The solid line in Fig.~\ref{Fourier} is a byproduct of the fit of the asymmetry spectrum. Tables~\ref{parameters_magnetic} and \ref{parameters_other} display all the numerical parameters of the model.  Notice that the two angles $\theta$ and $\varphi$ in Tab.~\ref{parameters_magnetic} have values corresponding to magnetic moments rotating in planes perpendicular to local three-fold axes. We found that any deviation from these values leads the model to substantially depart from the data. Therefore, in the final fit, these two values were fixed. 
Similarly we found that any small change in the muon position from the value $x_{\mu^+} = 0.532$ derived from the aforementioned paramagnetic phase data results in a worse fit of the data. 

\begin{table}
\begin{tabular}{ccccccc}
\hline\hline
$k$ & $m$ & $\theta$ & $\varphi$ & $\phi$ & $\xi$   \\
nm$^{-1}$ & $\mu_{\rm B}$ & deg. & deg. & deg. & nm \\ \hline
$0.35$ & $0.385\,(1)$ & $54.7$ & $45$ & $-2.04\,(11)$ & $258\,(35)$\\
\hline\hline
\end{tabular}
\caption{Parameters found for the magnetic structure of MnSi. The three rows correspond to the parameters defined in the main text, their units and their values with uncertainties, respectively. When no uncertainty is provided, it means that the parameter was fixed in the final fitting procedure.}
\label{parameters_magnetic}
\end{table}
\begin{table}
\begin{tabular}{cccc}
\hline\hline
$a_0$ & $r_\mu H/4\pi$ & $\lambda_Z$ & $\Delta_{\rm N}$\\
(-) & (-) & $\mu$s$^{-1}$ & mT \\ \hline
$0.250\,(3)$ & $-1.04\,(1)$ & 0.020\,(2) & $1.11\, (4)$ \\
\hline\hline
\end{tabular}
\caption{Parameters other than those shown in Table~\ref{parameters_magnetic} used in the fitting procedure.}
\label{parameters_other}
\end{table}

Interestingly, assuming $\lambda_Z$ to be independent of the orientation of the $Z$-axis, i.e., the initial muon polarization, we have numerically checked that $P_Z(t)$ is independent of this orientation. This means that the same $D_{\rm osc} (B)$ would have been measured for an other orientation of the initial muon polarization in the crystal lattice or in a powder sample. 

\section{Discussion}
\label{Discussion}

\begin{figure}
\centering
\begin{picture}(240,195)
\put(0,180){(a)}
\put(-1.5,70){\includegraphics{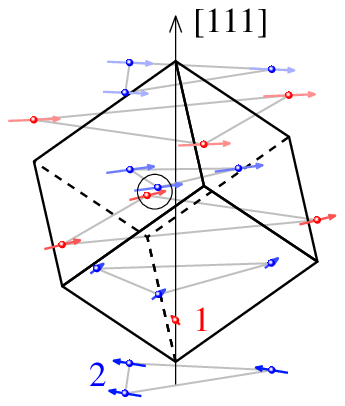}}
\put(120,180){(b)}
\put(118.5,70){\includegraphics{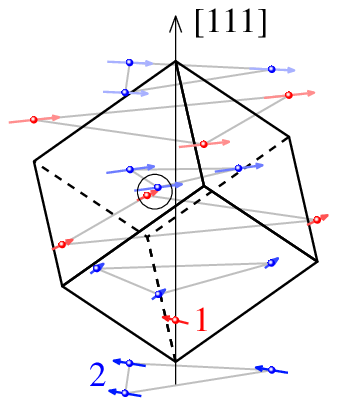}}
\put(0,52){(c)}
\put(0,0){\includegraphics[scale=1]{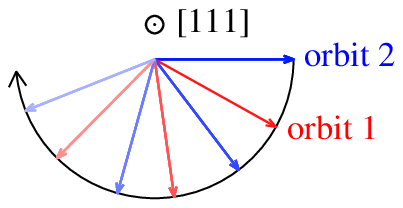}}
\put(120,52){(d)}
\put(120,0){\includegraphics[scale=1]{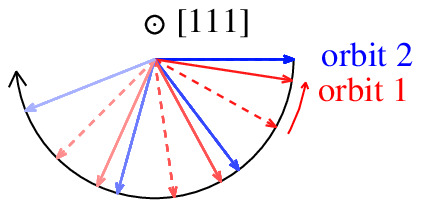}}
\end{picture}
\caption{(Color online) Schematic view of the MnSi magnetic structure. We consider domain A: the propagation vector is parallel to the cubic [111] axis. The structure can be viewed as a stacking of ferromagnetic Mn planes perpendicular to the [111] direction. Each plane consists of Mn atoms belonging exclusively either to orbit 1 or to orbit 2. The corresponding planes are alternatively stacked. The arrows represent the magnetic moments of the atoms: the same color is used for atoms belonging to a given plane. (a) Perpespective view of the currently accepted structure, as given by Eq.~\ref{moment_neutron}, and (b) of the magnetic structure deduced from the current study. Instead of being approximately oriented half-way between two orbit 2 planes, the orientation of the moments in orbit 1 is close to that of the plane situated below it. The (small) difference between the two configurations is most easily seen by comparing the relative orientation of the moments enlightened by the circle. A more spectacular view is provided by (c) and (d), corresponding respectively to (a) and (b), where the moments are projected in a plane perpendicular to [111]. Looking head on, the rotation is clockwise when moving along the positive direction, in accord with the magnetic structure chirality. In (d) the original orientation (dashed line) of the moments in orbit 1 is shown together with the orientation deduced from the present study (full line). 
For the sake of clarity the angular variations are amplified by a factor 10 in all panels, and the cubic unit cell is drawn in (a) and (b).}
\label{structure}
\end{figure}

We have successfully described the $\mu$SR asymmetry spectrum with the periodicity of the helix as given by neutron diffraction and a Mn magnetic moment consistent with the value extracted from magnetization measurements. The three characteristic fields are reproduced thanks to the finite $\phi$ value. The moments are still rotating in planes perpendicular to local three-fold axes. These are the main results of this work. In Fig.~\ref{structure} a picture of the magnetic structure in domain A is provided. While the $\phi$ value seems small at first sight, it has a noticeable effect on the magnetic structure: the moments of orbit 1 align nearly ferromagnetically with the moments belonging to the second-nearest Mn plane (orbit 2) --- recall that the Mn planes are not equidistant. In fact, the rotation of the Mn moments in these planes is almost locked. 

The large coherence length is a signature of a well ordered magnetic structure in our sample. The nuclear damping is large relative to the value $\Delta_{\rm N} = 0.36$~mT computed in the Van Vleck high-field limit.\cite{vanVleck48,Yaouanc11} However, intrinsic and muon-induced electric-field gradients acting on the $^{55}$Mn isotope\footnote{There is only one isotope for Mn and $^{29}$Si, the only non spin-less isotope of Si, has a spin 1/2} are expected to increase $\Delta_{\rm N}$.\cite{Hartmann77}  It is known that these gradients are strong as fingerprinted by the easily observed avoided-level-crossing resonances.\cite{Kadono93} A description of these electric-field gradient is still missing. It may explain the observed misfit on the left-hand side wing of the main peak; see Fig.~\ref{Fourier}. 

The hyperfine field parameter $H$ deduced from the analysis of spectra recorded in the paramagnetic state in Ref.~\onlinecite{Amato14} and here at 5~K are very close. This implies that we do not observe any rebuilding of the Fermi surface when crossing the magnetic phase transition. In addition, this consistency is an argument against any muon-induced magnetic structure change. The parameter $r_\mu H/4\pi$ is negative as always found. It is somewhat larger in absolute value than for metallic rare-earth or elemental ferromagnets.\cite{Yaouanc11} This may not be surprising given the correlated nature of the conduction electrons in MnSi.

The $\lambda_Z$ value measured here is much smaller than reported in Ref.~\onlinecite{Yaouanc05} for the same sample. This is obvious looking directly at the asymmetry spectrum, i.e., without any fitting. The origin of the discrepancy is unknown, but the previously published value was extracted from spectra recorded in a small longitudinal field (5~mT). We note that an unexplained rise of $\lambda_Z$ under the application of a small field has already been reported for other magnetic compounds.\cite{Dalmas06,Yaouanc11a,Baker12,Yaouanc15}

Our analysis of the $\mu$SR spectrum has been performed assuming the most frequent handedness for the crystal structure. However the two enantiomers exist\cite{Grigoriev10} and interestingly it was shown that the crystalline structure handedness determines the chirality of the magnetic structure.\cite{Grigoriev10} Assuming our sample to crystallize in the other chirality, i.e., $x_{\rm Mn} = 1-0.138$ = 0.862, and accordingly changing the muon site to $x_{\mu^+} = 1-0.532$ = 0.468, we have checked that an identical field distribution and therefore the same $P_Z(t)$ as shown in Figs.~\ref{Fourier} and \ref{Spectrum} are obtained, only provided the magnetic chirality is changed.\footnote{Changing the magnetic chirality essentially interchanges the position of the lower cutoff and singular peaks in the field distribution.} Therefore our conclusion about the MnSi magnetic structure is independent of the crystal handedness.

The question which naturally arises is the detection of the deviation from the originally inferred helical structure in neutron scattering measurements. In a usual experiment with unpolarized neutrons, the scattered intensity is proportional to the square modulus of the magnetic structure factor component perpendicular to the scattering vector. Recalling the magnetic structure factor definition 
\begin{eqnarray}
{\bf F}^{\rm mag}_{\ell}({\bf Q}) = \sum_{{\bf i},\gamma} {\bf m}_{\ell,i+d_\gamma}\exp\left[i{\bf Q}\cdot({\bf i}+{\bf d}_\gamma)\right], 
\label{structure_factor}
\end{eqnarray}
and using Eqs.~\ref{Moment_Fourier}, \ref{Moment_Fourier_component}, and \ref{Moment_Fourier_component_expression} we have
\begin{eqnarray}
{\bf F}^{\rm mag}_{\ell}({\bf k_\ell}) & = & \sum_{\gamma} {\bf m}_{\ell,d_\gamma,-{\bf k}_\ell} \label{structure_factor_MnSi}\\
& = & \frac{m\sqrt{n_c}}{2} \left[ 3+\exp(-i\phi)\right] \left( {\bf a}_{\ell,d_{\rm I}} -i\, {\bf b}_{\ell,d_{\rm I}}\right),
\nonumber 
\end{eqnarray}
since we have found the vectors ${\bf a}_{\ell,d_\gamma}$ and ${\bf b}_{\ell,d_\gamma}$ to be independent of $\gamma$. From $\phi$ = 0 to $-2.04^\circ$, the relative change in $\left| {\bf F}^{\rm mag}_{\ell}({\bf k_\ell}) \right|^2$ is about $3\times 10^{-4}$, making the magnetic structure derived from $\mu$SR very difficult to evidence in a neutron scattering measurement. 

As mentioned in Sec.~\ref{MnSi_basic}, Eq.~\ref{moment_neutron} is derived in the long wavelength limit. Starting from the same physical picture, but performing a microscopic analysis of the interaction energies, a deviation of the magnetic structure from a simple helix has been predicted\cite{Chizhikov12} with signatures reminiscent of those obtained from our representation theory analysis: a dephasing of the magnetic moments in the two orbits and a rotation plane not perpendicular to ${\bf k}_\ell$ for the moments of the atoms in orbit 2. We observed the first signature but not the second. Further studies of Hamiltonian models with an extension of the exchange interaction beyond nearest neighbours could be considered.\cite{Chizhikov13} Since the phase shift between certain subsequent planes is almost locked, we already infer the following consequence. To the interactions generally considered for the description of the magnetic properties of MnSi, namely the ferromagnetic exchange, DM, and weak anisotropic exchange interactions, listed here by order of decreasing magnitude, an additional term must contribute to the spin Hamiltonian. Since the presence of this term strongly decreases the effect of the DM interaction on the Mn atoms in one of the orbits it is not negligible.

When submitted to a relatively modest magnetic field, MnSi exhibits a so-called magnetic skyrmion lattice phase.\cite{Muhlbauer09a} This phase which is observed in the vicinity of the paramagnetic phase has been found in several other systems. While it occupies a small pocket of the temperature-magnetic field phase diagram of bulk materials, its stability range is dramatically enhanced in thin films,\cite{*[{See, e.g., }] [{}] Yu10a} suggesting the use of magnetic skyrmions in spintronics applications.\cite{Fert13} The skyrmion phase being described as a superposition of three helices akin to the ZF structure of MnSi, our result might have implications on the detailed arrangement of the spins in this phase and its interpretation in terms of microscopic interactions. Signatures of the phase shift could be found in the phason excitations predicted for a skyrmion lattice.\cite{Tatara14}

\section{Conclusions}
\label{Conclusions}

We have managed to analyze a zero-field $\mu$SR spectrum of MnSi using a conventional method dealing with the reciprocal space, combined with representation theory as applied to the determination of a magnetic structure. We find the phase of the moments in orbit 1 to be shifted by about $-2$ degrees relative to the value given by the scalar product ${\bf k}\cdot{\bf r}$.
The neutron diffraction results do not contradict our finding. However, given the small shift found for MnSi, it is a challenge to confirm it using a scattering method. Assuming such a phase slip to exist in other helimagnets, it might be easier to evidence it in a system with a shorter helix pitch.

Remarkably, there is no need to introduce exotic physics such as a muon-induced effect to understand the measured asymmetry spectrum. Only time-honored physics is required.  In addition to a new limit set on the knowledge of the magnetic structure of MnSi, this study provides a novel framework for the detailed refinement of subtle spin textures from $\mu$SR data.

\begin{acknowledgments}
We are grateful to P.J. Brown for an enlightening discussion about the s-domains in an incommensurate magnetic structure.
D.A. acknowledges partial financial support from UEFISCDI Project No. PN-II-ID-PCE-2011-3-0583 (85/2011), Romania.
\end{acknowledgments}

\appendix

\section{Magnetic structure of M\lowercase{n}S\lowercase{i} as determined by representation theory}
\label{Group_theory_application}

In the same way as vibrations in crystals are classified according to normal modes determined by representation theory,\cite{Tinkham64} magnetic structures compatible with crystal symmetry can be inferred with the help of the same theory; see Ref.~\onlinecite{Bertaut81} and references therein. 

The magnetic structures consistent with the crystal symmetries of MnSi have been determined using available codes. In particular, the basis vectors for the symmetry-allowed magnetic structures have been calculated with BasIREPS (FullProf)\cite{Rodriguez93} and SARA$h$ (Ref.~\onlinecite{Wills00}) which give consistent results.

Here we shall explain the method focusing on the A domain with ${\bf k}_{\rm A} = k(1,1,1)/\sqrt{3}$. Results for the other three k-domains can be obtained by simple symmetry arguments; a few hints will be given at the end of the section.
 
For the G$_{{\bf k}_{\rm A}}$ group which is made up of symmetry elements that do not change ${\bf k}_{\rm A}$ we find the Mn sites to split into two so-called crystallographic orbits. The first orbit only contains Mn$_{\rm I}$. The other three Mn atoms are in the second orbit. This classification can be physically understood since only for Mn$_{\rm I}$ is the [111] axis going through that position and is parallel to ${\bf k}_{\rm A}$; see Fig.~\ref{crystal_structure}.

For the first orbit we find three one-dimensional irreducible representations (irreps). Only one of them corresponds to a left-handed helix. Because our aim is to find a solution closely related with the currently accepted magnetic structure, we select this irrep. Introducing two unit vectors ${\bf a}_{{\rm A}, d_{\rm I}}$ and ${\bf b}_{{\rm A}, d_{\rm I}}$, perpendicular to each other and to $\hat{\bf k}_{\rm A}\equiv {\bf k}_{\rm A}/k_{\rm A}$ such that (${\bf a}_{{\rm A}, d_{\rm I}}$, ${\bf b}_{{\rm A}, d_{\rm I}}$, $\hat{\bf k}_{\rm A}$) is a right-handed basis, we write
\begin{eqnarray}
{\bf m}_{{\rm A},i + d_{\rm I}} & = &  m \left\{ \cos \left[{\bf k}_{\rm A}\cdot ({\bf i + d_{\rm I}}) + \alpha_{{\rm A},d_{\rm I}} \right] {\bf a}_{{\rm A}, d_{\rm I}} \right.  \cr
& & - \left. \sin \left [{\bf k}_{\rm A}\cdot ({\bf i + d_{\rm I}}) + \alpha_{{\rm A},d_{\rm I}} \right ] {\bf b}_{{\rm A},d_{\rm I}}\right\}
\label{moment_muon_A_I_general}
\end{eqnarray}
For definiteness, we will set 
\begin{eqnarray}
{\bf a}_{{\rm A}, d_{\rm I}} = 2^{-1/2}(1,\bar{1},0),\hspace{5mm}
{\bf b}_{{\rm A}, d_{\rm I}} = 6^{-1/2}(1,1,\bar{2}), 
\label{moment_muon_A_I_specific}
\end{eqnarray}
which does not impose restrictions to the set of allowed structures. Symmetry has nothing to say about $\alpha_{{\rm A},d_{\rm I}}$. The shift $\phi$ of this phase relative to the phase of moments in orbit 2 is of the uttermost importance for the description of the ZF spectrum. Remarkably, at position ${\rm I}$ the helix is as predicted by the Ginzbug-Landau expansion, since Eq.~\ref{moment_muon_A_I_general} corresponds {\em mutatis mutandis} to Eq.~\ref{moment_neutron} with $\alpha_0$ = $\alpha_{{\rm A},d_{\rm I}}$.

For the second orbit we find three one-dimensional irreps. Let us first consider the results for one of them: 
\begin{eqnarray}
{\bf m}_{{\rm A},i + d_{\rm II}} & = &  \cos \left[{\bf k}_{\rm A}\cdot ({\bf i + d_{\rm II}}) \right] {\bf U}_{{\rm A}, d_{\rm II}}   \cr
& - & \sin \left [{\bf k}_{\rm A}\cdot ({\bf i + d_{\rm II}}) \right ] {\bf V}_{{\rm A},d_{\rm II}}, \cr
{\bf m}_{{\rm A},i + d_{\rm III}} & = &  \cos \left[{\bf k}_{\rm A}\cdot ({\bf i + d_{\rm III}}) -2\pi/3 \right] {\bf U}_{{\rm A}, d_{\rm III}}   \cr
& - & \sin \left [{\bf k}_{\rm A}\cdot ({\bf i + d_{\rm III}}) -2\pi/3 \right ] {\bf V}_{{\rm A},d_{\rm III}},\cr
{\bf m}_{{\rm A},i + d_{\rm IV}} & = &  \cos \left[{\bf k}_{\rm A}\cdot ({\bf i + d_{\rm IV}}) -4\pi/3 \right] {\bf U}_{{\rm A}, d_{\rm IV}}   \cr
& - & \sin \left [{\bf k}_{\rm A}\cdot ({\bf i + d_{\rm IV}}) -4\pi/3 \right ] {\bf V}_{{\rm A},d_{\rm IV}}.
\label{moment_muon_A_II_general}
\end{eqnarray}
We have introduced 6 vectors: ${\bf U}_{{\rm A},d_\gamma}$ and ${\bf V}_{{\rm A},d_\gamma}$ for $\gamma \in \{ {\rm II},{\rm III},{\rm IV}\}$. They are written 
\begin{eqnarray}
{\bf U}_{{\rm A}, d_{\rm II}} & = & (u_1,u_2,u_3), \, \, {\bf V}_{{\rm A},d_{\rm II}} = (v_1,v_2,v_3), \cr
{\bf U}_{{\rm A}, d_{\rm III}} & = & (u_2,u_3,u_1), \, \, {\bf V}_{{\rm A},d_{\rm III}} = (v_2,v_3,v_1), \cr
{\bf U}_{{\rm A}, d_{\rm IV}} & = & (u_3,u_1,u_2), \, \, {\bf V}_{{\rm A},d_{\rm IV}} = (v_3,v_1,v_2),
\label{moment_muon_definitions}
\end{eqnarray}
and therefore depend only on six real numbers $u_i$ and $v_i$, for $i\in \{1,2,3\}$. The method to derive these components is explained in Appendix~\ref{Group_theory_basic}. 
In fact, the ${\bf U}_{{\rm A}, d_\gamma}$ (${\bf V}_{{\rm A}, d_\gamma}$) vector from a line in Eq.~\ref{moment_muon_definitions} is obtained from the corresponding vector of the previous line after rotation $R_{{\bf k}_{\rm A}}(4\pi/3)$ of angle $4\pi/3$ around ${\bf k}_{\rm A}$: for instance, ${\bf U}_{{\rm A}, d_{\rm IV}}$ is the image of ${\bf U}_{{\rm A}, d_{\rm III}}$ through rotation $R_{{\bf k}_{\rm A}}(4\pi/3)$.

Sticking to our guideline we impose the same $m$ value for the magnetic moment at the four different Mn sites in the crystal. Keeping with our previous notations we therefore set 
\begin{eqnarray}
{\bf U}_{{\rm A}, d_\gamma} & = & m\, {\bf a}_{{\rm A}, d_\gamma}, \ \ 
{\bf V}_{{\rm A},d_\gamma}  =  m\, {\bf b}_{{\rm A}, d_\gamma},
\label{moment_muon_A_basic_vectors_def}
\end{eqnarray}
for $\gamma$ = II, III, and IV, where ${\bf a}_{{\rm A}, d_\gamma}$ and ${\bf b}_{{\rm A}, d_\gamma}$ are two orthogonal unit vectors. Contrary to the first orbit (Eq.~\ref{moment_muon_A_I_general}) and to the Ginzburg-Landau derivation (Eq.~\ref{moment_neutron}), the vector ${\bf n}_{{\rm A},d_{\gamma}} \equiv {\bf a}_{{\rm A}, d_\gamma} \times {\bf b}_{{\rm A}, d_\gamma}$ is not necessarily collinear to ${\bf k}_{\rm A}$. Still, when it is collinear it is easily shown that the magnetic structure described by Eq.~\ref{moment_neutron} can be obtained from Eqs.~\ref{moment_muon_A_II_general} and \ref{moment_muon_A_basic_vectors_def} after appropriately setting ${\bf a}_{{\rm A}, d_\gamma}$ and ${\bf b}_{{\rm A}, d_\gamma}$ in the plane defined by ${\bf a}_{\ell={\rm A}}$ and ${\bf b}_{\ell={\rm A}}$.

To proceed we need to specify the three Euler angles defining the orthonormal basis $({\bf a}_{{\rm A}, d_{\rm II}}, {\bf b}_{{\rm A}, d_{\rm II}}, {\bf n}_{{\rm A},d_{\rm II}})$ in the crystallographic cubic axes. In line with the restriction we prescribe for the possible magnetic structures, we will set the third Euler angle defining the orientation of ${\bf a}_{{\rm A} ,_{\rm II}}$ and ${\bf b}_{{\rm A}, d_{\rm II}}$ in the plane perpendicular to ${\bf n}_{{\rm A},d_{\rm II}}$ to the value imposed by Eq.~\ref{moment_neutron}. Then the remaining free parameters are the polar and azimuthal angles $\theta$ and $\varphi$ for ${\bf n}_{{\rm A},d_{\rm II}}$. We have
\begin{eqnarray}
{\bf  a}_{{\rm A}, d_{\rm II}} & = & (\sin \varphi, - \cos \varphi, 0),\cr
{\bf  b}_{{\rm A}, d_{\rm II}} & = &( \cos \varphi \cos \theta, \sin \varphi \cos \theta, -\sin \theta).
\label{moment_muon_A_basic_vectors}
\end{eqnarray}
The basis relative to Mn sites ${\rm III}$ and ${\rm IV}$ is then directly obtained from Eqs.~\ref{moment_muon_definitions} and \ref{moment_muon_A_basic_vectors_def}. When $\varphi = \varphi_0 = 45^\circ$ and $\theta = \theta_0 = 54.7^\circ$ [more precisely $\cos\theta_0 = (1/3)^{1/2}$ and $\sin\theta_0 = (2/3)^{1/2}$], and $\phi \equiv \alpha_{{\rm A},d_{\rm I}}$ = 0, we recover the known magnetic structure.

We do not consider the other irreps of the second orbit since, in the limit of vectors ${\bf a}_{{\rm A}, d_\gamma}$ and ${\bf b}_{{\rm A}, d_\gamma}$ perpendicular to ${\bf k}_{\rm A}$ they lead to a dephasing of $\pm 2\pi/3$ of the moments at the three Mn sites, a solution at strong variance from Eq.~\ref{moment_neutron}. 

In summary, while limiting ourselves to a magnetic structure closely related to the currently accepted one, the application of representation theory provides us with two directions for relaxing the structure given by Eq.~\ref{moment_neutron}. The first one is the absence of link between the phase of the magnetic moment at site ${\rm I}$ with that at sites ${\rm II}$, ${\rm III}$, and ${\rm IV}$.
The second is the fact that the moments at sites ${\rm II}$, ${\rm III}$, and ${\rm IV}$ might not rotate in a plane perpendicular to ${\bf k}_{\rm A}$.
The spectrum refinement is made allowing the three angles to vary from their initial values $\varphi = \varphi_0 =   45^\circ$, $\theta = \theta_0 =   54.7^\circ$ and $\phi= \phi_0 = 0$.

A similar inference can be carried out for the K-domains B, C, and D. For example, for domain B with ${\bf k}_{\rm B}  =  k ({\bar 1},{\bar 1},1)/\sqrt {3}$ site Mn$_{\rm II}$ belongs to the first orbit and Mn$_{\rm III}$, Mn$_{\rm IV}$ and Mn$_{\rm I}$ to the second orbit. Representation theory allows to derive equations similar to Eqs.~\ref{moment_muon_A_I_general} and \ref{moment_muon_A_II_general} for the two orbits and Eq.~\ref{moment_muon_definitions} must be adapted for a rotation around ${\bf k}_{\rm B}$. Restricting the possible magnetic structures as for domain A, equations similar to Eq.~\ref{moment_muon_A_basic_vectors} are obtained. And so on for domains C and D.

\section{Physics of the second orbit}
\label{Group_theory_basic}

Here we discuss the physics of the second orbit. The natural representation $\Gamma$ can be decomposed in terms of the weighted sum of three one-dimensional irreps such that $\Gamma = 3 \Gamma^1 \oplus 3 \Gamma^2 \oplus 3 \Gamma^3$. From the results of the two available computer codes we have written down three systems of linear equations for the symmetry-allowed basis functions for domain A. The three functions $F_{1,i}$ for $\Gamma^1$ in terms of the spin coordinates $S^\alpha_{{\rm A},d_\gamma}$, with $\alpha = \{x,y,z \}$, follow the system of three linear equations:
\begin{eqnarray}
F_{1,1} & = &  S^x_{{\rm A},d_{\rm II}} + S^y_{{\rm A},d_{\rm III}} +S^z_{{\rm A},d_{\rm IV}} ,\cr
F_{1,2} & = &  S^y_{{\rm A},d_{\rm II}} + S^z_{{\rm A},d_{\rm III}} +S^x_{{\rm A},d_{\rm IV}} ,\cr  
F_{1,3} & = &   S^z_{{\rm A},d_{\rm II}} + S^x_{{\rm A},d_{\rm III}} +S^y_{{\rm A},d_{\rm IV}} . 
\label{Group_theory_basic_1}
\end{eqnarray}
In the same way, for $\Gamma^2$ we derive
\begin{eqnarray}
F_{2,1} & = &  S^x_{{\rm A},d_{\rm II}} + a S^y_{{\rm A},d_{\rm III}} +a^* S^z_{{\rm A},d_{\rm IV}} ,\cr
F_{2,2} & = &  S^y_{{\rm A},d_{\rm II}} + a S^z_{{\rm A},d_{\rm III}} +a^* S^x_{{\rm A},d_{\rm IV}} ,\cr  
F_{2,3} & = &   S^z_{{\rm A},d_{\rm II}} + a S^x_{{\rm A},d_{\rm III}} + a^* S^y_{{\rm A},d_{\rm IV}},
\label{Group_theory_basic_2}
\end{eqnarray}
and finally for $\Gamma^3$ we get
\begin{eqnarray}
F_{3,1} & = &  S^x_{{\rm A},d_{\rm II}} + a^* S^y_{{\rm A},d_{\rm III}} +a S^z_{{\rm A},d_{\rm IV}} ,\cr
F_{3,2} & = &  S^y_{{\rm A},d_{\rm II}} + a^* S^z_{{\rm A},d_{\rm III}} +a S^x_{{\rm A},d_{\rm IV}} ,\cr  
F_{3,3} & = &   S^z_{{\rm A},d_{\rm II}} + a^* S^x_{{\rm A},d_{\rm III}} + a S^y_{{\rm A},d_{\rm IV}}.
\label{Group_theory_basic_3}
\end{eqnarray}
Here we have introduced the phase factor $a = - \exp(i \pi/3) = \exp(i 4\pi/3) $. 

If MnSi orders magnetically, for example, according to the first irrep, the coordinates for the second and third irreps vanish, i.e., $F_{2,i} = F_{3,i} =0$.\cite{Bertaut81} This leads to the determination of the components of ${\bf S}_{{\rm A},d_\gamma}$ in terms of unknown constants. This rule is obviously valid for the three irreps. Equations~\ref{moment_muon_A_II_general} and \ref{moment_muon_definitions} in Appendix~\ref{Group_theory_application} are derived in this way.
By definition, ${\bf S}_{{\rm A},d_\gamma} = {\bf U}_{{\rm A},d_\gamma} + i {\bf V}_{{\rm A},d_\gamma}$.

\bibliography{reference}

\end{document}